\documentclass[aps,pre,citeautoscript,preprint,citeautoscript,showkeys]{revtex4-2}
  \usepackage[main=english, catalan, english]{babel}
\DeclareUnicodeCharacter{2212}{-}
\usepackage{graphicx}
\usepackage{amsmath}

\date{\today}
\title{}
\begin{document}

\title{One Component Plasmas Frequency Moments from the ISF}
\author{D. Ballester}
\affiliation{Al-Farabi Kazakh National University, 050040 Almaty, Kazakhstan}


\date{\today}

\begin{abstract}
Dynamic structural characteristics of
classical one component plasmas (OCP) are an essential tool for 
understanding of strongly coupled plasmas through experiments and numerical simulations.
Here, instead of the dynamic structure factor, we focus our attention
on the intermediate scattering function (ISF). Theoretical results are compared to those of obtained by the
method of molecular dynamics (MD).
\end{abstract}


\maketitle
\section{Introduction}
\label{sec:org540e29a}

The One Component Plasma (OCP) model has played a prominent role in
the field of plasma physics. This model can be thought as the classical equivalent of the electron gas \cite{bohm-1949-theor-plasm-oscil}. It considers a plasma consisting
of only one type of charged particles, which is embedded in a uniform
neutralizing background. The properties of the system are completely
defined by the coupling coefficient \(\Gamma = q^2/(akBT)\),
where \(q\) denotes the charge, \(a = (3/4\pi n)^{1/3}\) is the Wigner-Seitz
radius, \(n\) is the density of charged particles, and \(T\) is the
temperature \cite{brush-1966-monte-carlo}  

Of particular interest is the strongly coupled regime, \(\Gamma \le  1\), where perturbative approximations break down. Some application include astrophysical plasmas, such as, compact degenerate stars, white dwarfs interiors and neutron stars \cite{chabrier-1992-white-dwarf,ostriker-1969-natur-pulsar,lamb-1975-evolut-cryst,kugler-1969-anhar-contr,chabrier-1993-quant-effec,baiko-2021-i-ab}.

Thermodynamic properties of the OCP, pair correlations and the onset of 
crystallization of the system at a high
coupling were studied using Monte Carlo simulations \cite{brush-1966-monte-carlo,slattery-1980-improv-equat,stringfellow-1990-equat-state}

Equilibrium, transport, and dynamical characteristics of the OCP have been
investigated in a series of
Monte Carlo (MC) and Molecular Dynamics (MD) simulation studies \cite{hansen-1973-statis-mechan,pollock-1973-statis-mechan,hansen-1975-statis-mechan}. 

With the advent  of modern computational power, it has been possible to increase the accuracy \cite{donko-2002-cagin-partic,daligault-2006-liquid-state,daligault-2012-diffus-ionic,korolov-2015-dynam-struc}.
\section{Dynamic structure factor and frequency moments}
\label{sec:orgaf24f84}

The dynamic structure factor (DSF) \(S(\mathbf{q},\omega)\) constitutes the
central 
property in scattering experiments. It contains
dynamical information needed to properly describe and
model collective modes and transport \cite{komatsu-2004-large-scale}, while also
providing information on stopping power \cite{deutsch-1995-correl-stopp,deutsch-2019-correl-ion}, neutron scattering \cite{brandt-2009-dynam-struc}, and X-ray Thomson Scattering \cite{murillo-2010-x-ray}. 

It is given by the Fourier
transform 
of the density-density correlaction function
\cite{2013-theor-simpl-liquid},

\begin{equation}\label{DSF}
S(\mathbf{q},\omega) = \int_{-\infty}^\infty \textnormal{d}t\ e^{i\omega t} 
\langle \hat n(\mathbf{q},t)\hat n(-\mathbf{q},0) \rangle  _T \ ,
\end{equation}

Frequency moments of DSF can be derived as follows \cite{ichimaru-1975-sum-rule}

\begin{equation}\label{DSF-mom}
S_\nu (k) = \int_{-\infty}^\infty \omega^k  S (k,\omega)  \textnormal{d}\omega
\end{equation}

in a OCP we have (in terms of a dimensionless \(q = ka\)):

\begin{equation}\label{DSF-mom-S0}
S_0 (k) = S (q) 
\end{equation}

\begin{equation}\label{DSF-mom-S2}
S_2 (k) = \frac{q^2}{3\Gamma} \omega_p^2
\end{equation}

\begin{equation}\label{DSF-mom-S4}
S_4 (k) =   \frac{q^2}{3\Gamma} \omega_p^4 \left[ 1 + q^2 + I (q) \right] 
\end{equation}

where \cite{hansen-1973-statis-mechan,adamjan-1993-dynam-static}

\begin{equation}\label{DSF-mom-I}
I (k) = \frac{1}{3\pi}\int_{0}^\infty [S (p) - 1]
\left[ \frac56 - \frac{p^2}{2q^2} + 
\left(\frac{p^3}
{4q^3} - \frac{p}{4q} + \frac{q}{2p} \right)
\textnormal{ln} \left| \frac{q + p}{q - p} \right| \right]  p^2 \textnormal{d}p
\end{equation}

\(S (q)\) is the static structure factor and \(\omega_p = \sqrt{4\pi ne^2/m}\) is the plasma frequency.
\section{Intermediate Scattering Function}
\label{sec:orgaffb113}

The intermediate scattering function \(F(\mathbf{q},t)\) being defined as \cite{hansen-1973-statis-mechan,2013-theor-simpl-liquid},

\begin{equation}\label{ISF1}
F(\mathbf{q},t) = \langle \hat n(\mathbf{q},t)\hat n(-\mathbf{q},0) \rangle  _T\ .
\end{equation}

Clearly,  \(F(\mathbf{q},t)\) can be calculated as the inverse Fourier transform of the \(S(\mathbf{q},\omega)\) given Eq. \eqref{DSF}

\begin{equation}\label{ISF2}
F(\mathbf{q},t) = \int_{-\infty}^\infty \textnormal{d}t\ e^{- i\omega t} 
 S(\mathbf{q},\omega) 
\end{equation}

Next, consider the partial derivative of the with respect to time \eqref{ISF2}
\begin{equation} \label{deriv}
\frac{\partial^\nu}{\partial t^\nu}F(\mathbf{q},t)=(-i)^\nu\int\limits_{-\infty}^{\infty} d\omega\, \omega^\nu e^{- i\omega t}S(\mathbf{q},\omega)\ .
\end{equation}

In particular, the \$t-\$derivative at the origin is given by

\begin{equation} \label{derivative}
\left. \frac{\partial^\nu}{\partial t^\nu}F(\mathbf{q},t)\right|_{t=0} =  \left(-i\right)^\nu \int\limits_{-\infty}^{\infty} d\omega\, \omega^\nu S(\mathbf{q},\omega)\ ,
\end{equation}

Hence,  one can express \(F(\mathbf{q},t)\) around \(t=0\) as a Taylor expansion \cite{hansen-1973-statis-mechan,2013-theor-simpl-liquid},

\begin{eqnarray}
F(\mathbf{q},t) &=& \sum_{\nu=0}^\infty \left\{
\frac{1}{\nu!} \left. \frac{\partial^\nu F(\mathbf{q},t)}{\partial t^\nu}\right|_{t=0}t^\nu
\right\} \\
&=& \sum_{\nu=0}^\infty \frac{\left(-i\right)^\nu}{\nu!} S_\nu(\mathbf{q})t^\nu \ . \label{Taylor}
\end{eqnarray}

where we had made used of Eq. \eqref{DSF-mom}. Indeed, we could use these time derivatives to compute the DSF frequency moments as follows

\begin{eqnarray}
\left. \frac{\partial^\nu F(\mathbf{q},t)}{\partial t^\nu}\right|_{t=0}
= \left(-i\right)^\nu S_\nu(\mathbf{q}) . \label{Taylor2}
\end{eqnarray}

In practice, one truncates Eq. \eqref{Taylor} at some finite degree 
\(\nu_\textnormal{max}\). For a classical OCP, at least up to \(\nu_\textnormal{max}\leq6\) are known to converge, with odd DSF frequency moments being zero \cite{ichimaru-1975-sum-rule} 
\section{Numerical results}
\label{sec:org5cc1470}

\subsection{MD simulations using LAMMPS}
\label{sec:orge21d1ff}

LAMMPS is a general-purpose simulation platform used for MD
simulations. Here we simulate an OCP model plasma with Coulomb
potential, in a neutralizing homogeneous background, using the
Particle Particle Particle-Mesh (PPPM) algorithm \cite{Hockney_1988}
that makes it possible to fully account for the long-range Coulomb
interparticle potential.

In all simulations the number of particles is \(N=10000\).
\subsubsection{\(\Gamma=10\) OCP using LAMMPS MD.}
\label{sec:orgad5f4f6}

\begin{figure}[ht]
  \centering
  \includegraphics[width=9cm]{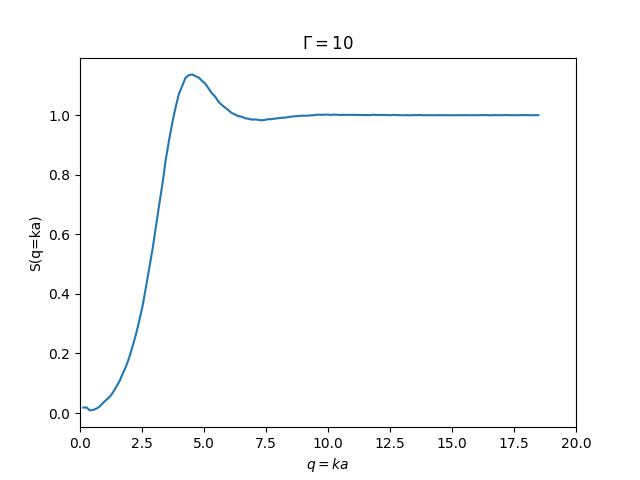}
  \caption{Static structure factor using LAMMPS and $\Gamma=10$.}  \label{fig:SqG10}
\end{figure}

The first MD simulation corresponds to \(\Gamma=10\). Fig. \ref{fig:SqG10}
shows the Static Structure Factor for such thermodynamic conditions.

\begin{figure}[ht]
  \centering
  \includegraphics[width=9cm]{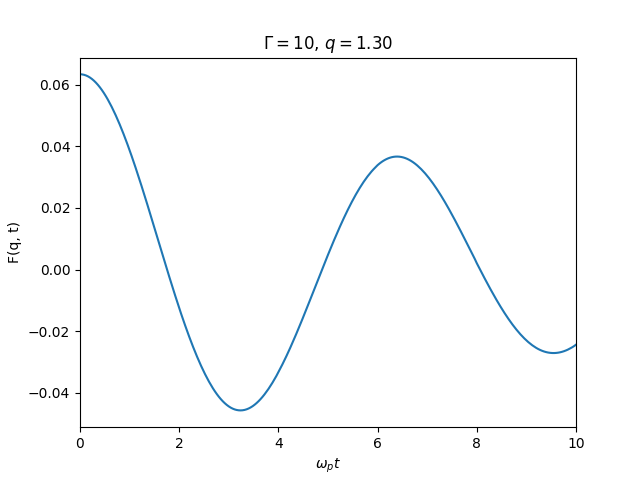}
  \caption{Intermediate Scattering Function using LAMMPS and $\Gamma=10$.}  \label{fig:FqtG10q24}
\end{figure}

\begin{figure}[ht]
  \centering
  \includegraphics[width=9cm]{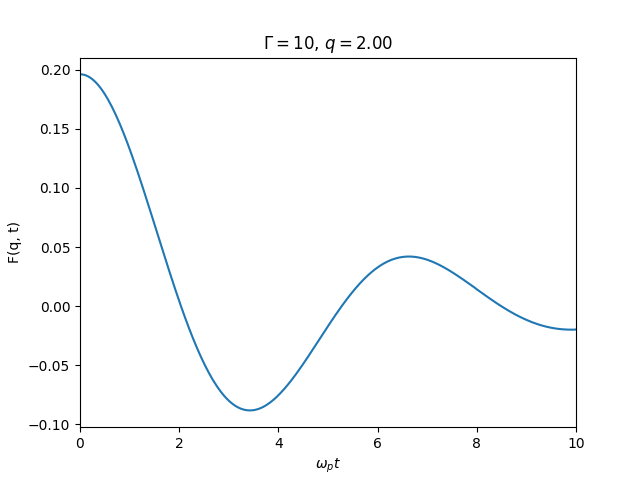}
  \caption{Intermediate Scattering Function using LAMMPS and $\Gamma=10$.}  \label{fig:FqtG10q37}
\end{figure}

Next, Figs. \ref{fig:FqtG10q24} and \ref{fig:FqtG10q37} depict the
Intermediate Scattering Function for two different values of \(q=ka\).

Following Eqs. \eqref{Taylor2} we can compute the frequency moments of
Dynamic Structure Factor as time derivatives of the Intermediate
Scattering Function, evaluated at \(t=0\), for different values of \(q\).

\begin{figure}[ht]
  \centering
  \includegraphics[width=9cm]{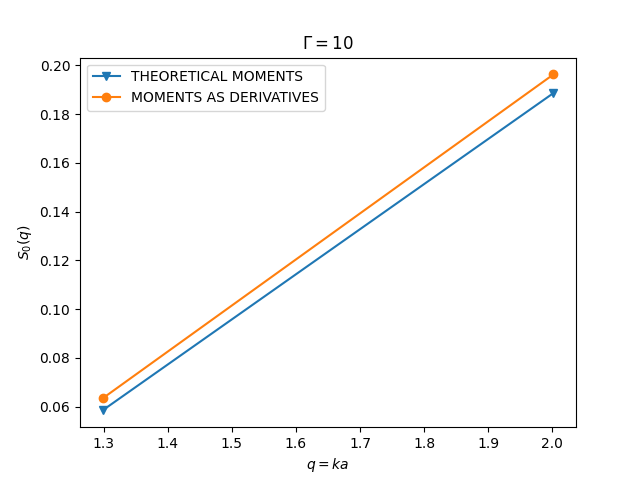}
  \caption{$S_0$ frequency moment using LAMMPS and $\Gamma=10$ from ISF.}  \label{fig:resultsG10S0LAMMPS}
\end{figure}

\begin{figure}[ht]
  \centering
  \includegraphics[width=9cm]{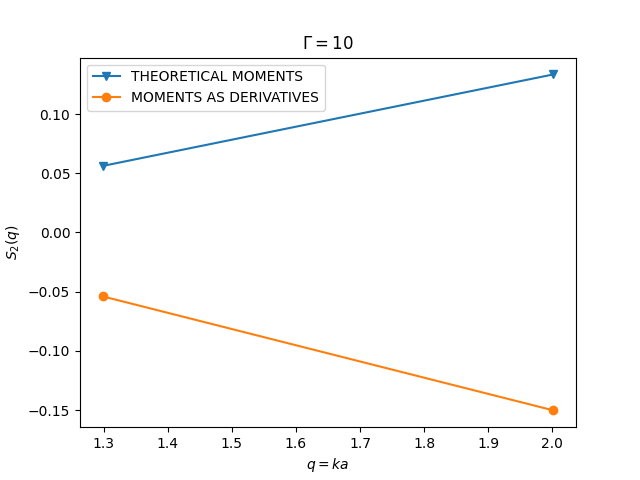}
  \caption{$S_2$ frequency moment using LAMMPS and $\Gamma=10$ from ISF.}  \label{fig:resultsG10S2LAMMPS}
\end{figure}

\begin{figure}[ht]
  \centering
  \includegraphics[width=9cm]{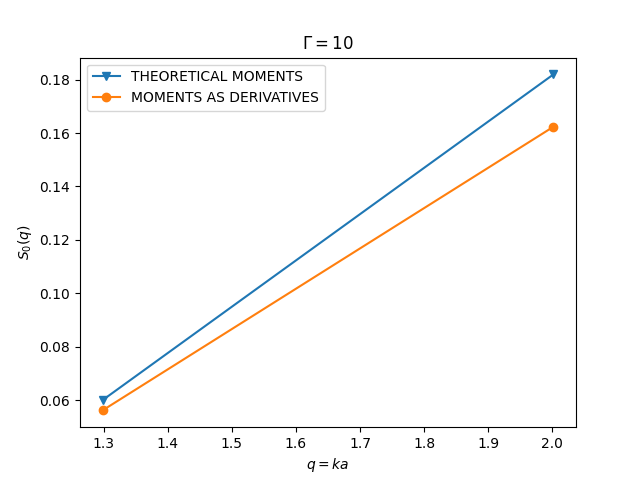}
  \caption{$S_4$ frequency moment using LAMMPS and $\Gamma=10$ from ISF.}  \label{fig:resultsG10S4LAMMPS}
\end{figure}

Indeed, Figs. \ref{fig:resultsG10S0LAMMPS}, \ref{fig:resultsG10S2LAMMPS} and
\ref{fig:resultsG10S4LAMMPS} correspond to the frequency moments \eqref{DSF-mom-S0},
\eqref{DSF-mom-S2} and \eqref{DSF-mom-S4}, respectively. Notice that the second
derivative for the second moment should be negative.

\subsubsection{\(\Gamma=70\) OCP using LAMMPS MD.}
\label{sec:orgb514ad3}

\begin{figure}[ht]
  \centering
  \includegraphics[width=9cm]{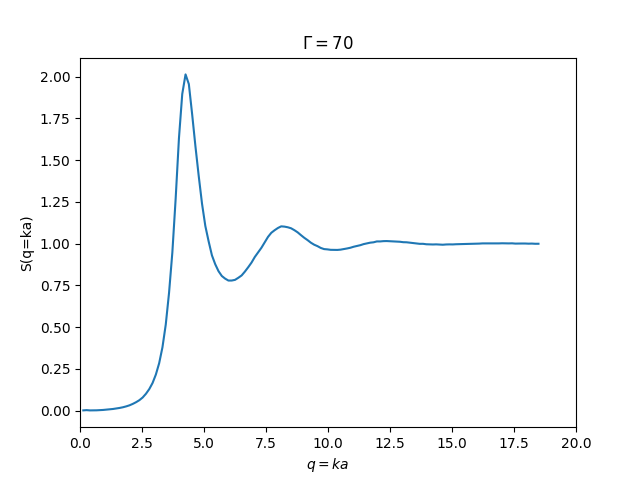}
  \caption{Static structure factor using LAMMPS and $\Gamma=70$.}  \label{fig:SqG70}
\end{figure}

The next MD simulation corresponds to \(\Gamma=70\). Fig. \ref{fig:SqG70}
shows the Static Structure Factor.

\begin{figure}[ht]
  \centering
  \includegraphics[width=9cm]{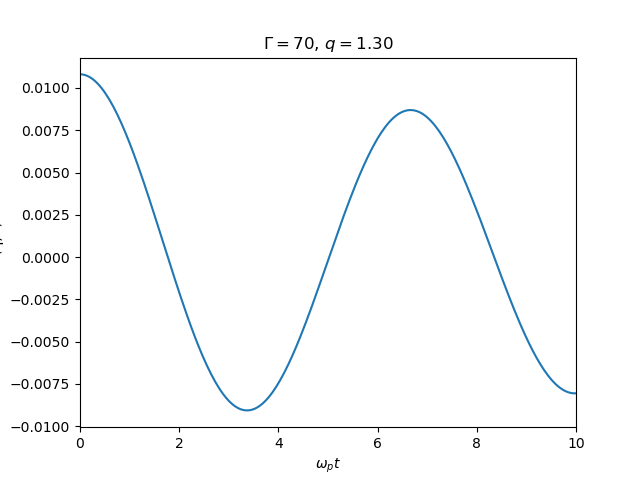}
  \caption{Intermediate Scattering Function using LAMMPS and $\Gamma=70$.}  \label{fig:FqtG70q24}
\end{figure}

\begin{figure}[ht]
  \centering
  \includegraphics[width=9cm]{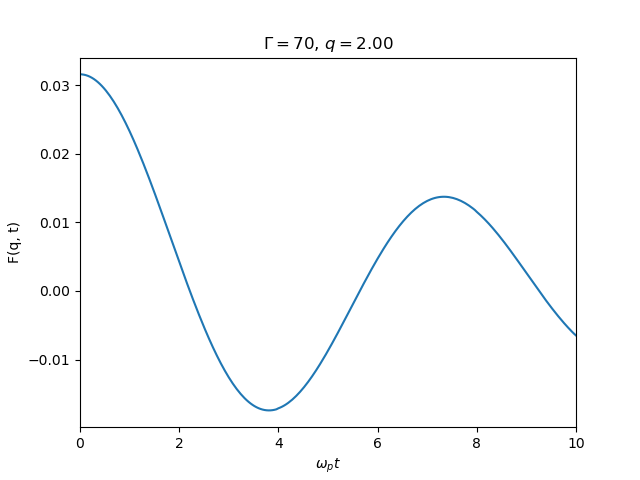}
  \caption{Intermediate Scattering Function using LAMMPS and $\Gamma=70$.}  \label{fig:FqtG70q37}
\end{figure}

Next, Figs. \ref{fig:FqtG70q24} and \ref{fig:FqtG70q37} depict the
Intermediate Scattering Function for two different values of \(q=ka\).

\begin{figure}[ht]
  \centering
  \includegraphics[width=9cm]{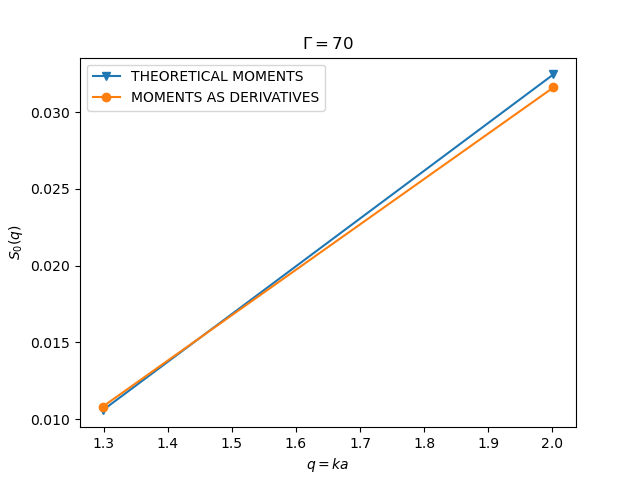}
  \caption{$S_0$ frequency moment using LAMMPS and $\Gamma=70$ from ISF.}  \label{fig:resultsG70S0LAMMPS}
\end{figure}

\begin{figure}[ht]
  \centering
  \includegraphics[width=9cm]{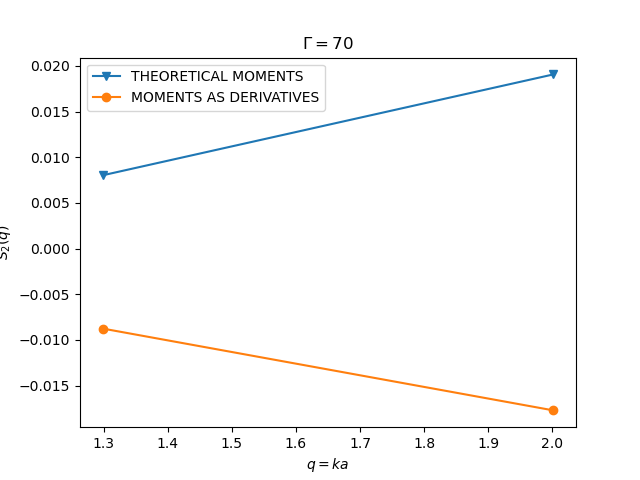}
  \caption{$S_2$ frequency moment using LAMMPS and $\Gamma=70$ from ISF.}  \label{fig:resultsG70S2LAMMPS}
\end{figure}

\begin{figure}[ht]
  \centering
  \includegraphics[width=9cm]{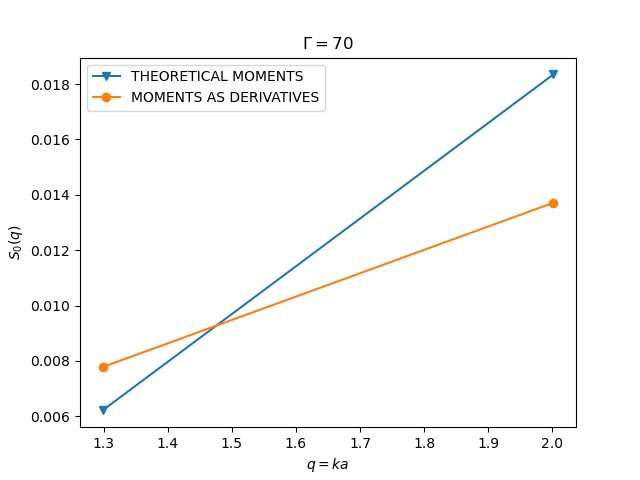}
  \caption{$S_4$ frequency moment using LAMMPS and $\Gamma=70$ from ISF.}  \label{fig:resultsG70S4LAMMPS}
\end{figure}

As before, Figs. \ref{fig:resultsG70S0LAMMPS}, \ref{fig:resultsG70S2LAMMPS} and
\ref{fig:resultsG70S4LAMMPS} correspond to the frequency moments \eqref{DSF-mom-S0},
\eqref{DSF-mom-S2} and \eqref{DSF-mom-S4}, respectively. Notice that the second
derivative for the second moment should be negative.

\section{Conclusions}
\label{sec:org7881942}

We have revisited the dynamic theory and simulation of classical strongly coupled plasmas, through the study of their known convergent sum-rules or frequency moments for different values of the coupling strength.

In particular, it is often thought that to interpret numerical results MD simulations, one must compute the DSF 

Here we calculate frequency moments from both, the DSF as integrals and, directly, from the ISF as time derivatives, for two different MD simulation codes. We find good agreement for all cases, when compared to the theoretical ones.





\bibliography{references}
\end{document}